\documentstyle[prb,aps,preprint]{revtex}

\begin{document}
\date{\today}
\draft

\title{Light scattering from a periodically modulated two dimensional
  electron gas with partially filled Landau levels}

\author{Arne~Brataas$^1$ and C. Zhang$^2$ and K. A. Chao$^{1,3}$}

\address{$^1$ Department of Physics, Norwegian University of Science
  and Technology, \\ N-7034 Trondheim, Norway.}

\address{$^2$ Department of Physics, University of Wollongong, New
  South Wales 2522, Australia.}

\address{$^3$ Department of Theoretical Physics, Lund University,
  S-223 26 Lund, Sweden.}

 \draft

\maketitle

\begin{abstract}
  We study light scattering from a periodically modulated two
  dimensional electron gas in a perpendicular magnetic field.  If a
  subband is partially filled, the imaginary part of the dielectric
  function as a function of frequency contains additional
  discontinuities to the case of completely filled subbands.  The
  positions of the discontinuities may be determined from the partial
  filling factor and the height of the discontinuity can be directly
  related to the modulation potential.  The light scattering cross
  section contains a new peak which is absent for integer filling.
\end{abstract}



\newpage

Since the Weiss oscillation was first observed in the
magneto-resistivity several years ago,\cite{Weiss89:179} a
considerable amount of work, both experimentally and theoretically,
have been carried out on the electronic and transport properties of
two dimensional electron system under a periodic potential and a
constant magnetic
field.\cite{Winkler89:1177,Gerhardts89:1173,Weiss89:13020,Pfannkuche92:12606,Gerhardts90:1473,Zhang90:12850,Cui89:2598,Zhang90:2207,Manolescu95:1703}
Most recent works include the DC transport in a strong anti-dot
system\cite{Weiss93:4118,Rotter96:4452} and the observation of quantum
fractal like energy spectrum.\cite{Schlosser96:683} Weiss oscillation
can be understood as a type of commensurability oscillation originated
from the interplay of two different length scales of the system, the
periodicity of the modulation potential $a$, and the radius of the
cyclotron motion $R_c$.  The DC resistivity has a set of minima
whenever the condition $2R_c=(n-1/4)a$ is satisfied, where $n$ is any
integer. Most of the current investigations have been limited to the
case where the external field has zero frequency. In a recent paper by
Stewart and Zhang,\cite{Stewart95:R17036} the dielectric response at
finite frequency and wave-numbers was calculated.  Their result
indicates that the modulation induced structure at finite frequency is
much richer than that of the static case. Most noticeably, the
electron-hole pair excitation contains a set of sub-singularities at
the excitation band edges.

In this brief report, we investigate the density response function
within the random-phase-approximation for a two dimensional electron
gas under a weak periodic modulation and a constant magnetic field. We
shall pay special attention to the case where the Fermi level lies
within a Landau band.  In what follows, we shall show that this
partially filled Landau band at the Fermi level has profound effect on
the dielectric response and in turn alters the light scattering cross
section from such a system. Our main results are: (i) Electron-hole
pair excitation has a step-like (discontinuous) behavior around
$\omega=n \omega_c$.  This step can be determined analytically for the
case of weak modulation.  Thus the amplitude of the modulation
potential, which so far has been rather difficult to measure
experimentally, can now be precisely determined.  (ii) Also around
$\omega=n \omega_c$, the real part of the dielectric function has a
logarithm divergence.  (iii) When the new effects of (i) and (ii) are
included in the density response function, a new sharp peak can be
observed in the light scattering cross section.

We consider a two dimensional electron gas where a static magnetic
field $B$ is perpendicular to the plane.  A weak periodic potential is
applied in the $x$-direction
\begin{equation}
  V(x) = V_0 \cos(K x) \, ,
\end{equation}
where $K=2\pi/a$ and $a$ is the period of the modulation.  In the
Landau gauge the single-particle wave functions are of the form
$\psi_{nk}(x,y) \propto \exp{(\imath k_y y)} \phi_{nx_{0}}(x)$ where
$x_0=k_y l^2$ (magnetic length, $l=[\hbar/(m^* \omega_c)]^{1/2}$) is
the center coordinate and $\phi_{nx_{0}}(x)$ is the eigenfunction of
the one-dimensional Hamiltonian
\begin{equation}
  H = - \frac{\hbar^2}{2 m^*} \frac{d^2}{dx^2} + \frac{1}{2} m^*
  \omega_c^2 (x-x_0)^2 + V_0 \cos{(K x)} \, ,
\end{equation}
where $m^*$ is the effective mass and $\omega_c = eB/(m^* c)$ is the
cyclotron frequency.  We assume that the modulation potential is a
weak perturbation, $V_0 \ll E_F$, where $E_F$ is the Fermi energy.
The energy spectrum to linear order in the modulation potential
is\cite{Gerhardts89:1173}
\begin{equation}
E_n(x_0) = \Bigl(n+ \frac{1}{2} \Bigr) \hbar \omega_c + U_n \cos{(K x_0)} \, ,
\end{equation}
where $U_n = V_0 \exp(-{\cal H}/2) L_n({\cal H})$, ${\cal H} =
(Kl)^2/2$ and $L_n({\cal H})$ is a Laguerre polynomial.

In order to simplify the discussion of the discontinuity in the
imaginary part of the dielectric function as a function of the
frequency let us consider only vertical transition.  The energy
difference for the transition from state $m^{\prime},x_0$ to state
$m^{\prime}+m,x_0$ is then ($m > 0$)
\begin{eqnarray}
  E_{m+m^{\prime}}(x_0) - E_{m^{\prime}}(x_0) & = &m \hbar \omega_c +
  \nonumber \\ & & (U_{m+ m^{\prime}}- U_{m^{\prime}}) \cos{(K x_0)}.
\end{eqnarray}
For sufficiently small ${\cal H}$ we may approximate $L_n({\cal H}) =
1 - n {\cal H}$, so that $U_{m+m^{\prime}}- U_{m} = - V_0 m {\cal H}
\exp(-{\cal H}/2)$.  We also have $U_n({\cal H}) > 0$ for small $n$
and ${\cal H}$.  This means that the energy difference between the
states $m+m^{\prime},x_0$ and $m^{\prime},x_0$ is less than $m \hbar
\omega_c$ for $ 0 < Kx_0 < \pi/2$ and $3\pi/2 < Kx_0 < 2\pi$.  On the
other hand the energy difference is larger than $m \hbar \omega_c$ for
$\pi/2 < Kx_0 < 3\pi/2$.  Let us now consider a situation where the
last Landau band (with index $n_F$) is half filled.  At $\omega \sim
\omega_c$, two transition processes ($n_F-1 \rightarrow n_F$ and $n_F
\rightarrow n_F+1$) contribute to the electron-hole pair excitation.
However the transition $(n_F-1\rightarrow n_F)$ only contributes for
$\omega < \omega_c$ while for $\omega > \omega_c$ only the transition
$(n_F \rightarrow n_F + 1)$ contributes (see Fig.\ (\ref{f:dis})).  At
$\omega=\omega_c$ the contributions from these two distinct
transitions are different and thus a discontinuity occurs at
$\omega=\omega_c$.  One can immediately generalize this conclusion
for excitations around $\omega \sim m\omega_c$ where a similar
discontinuity occurs due to the different contributions from the
$(n_F-m \rightarrow n_F)$ transition and the $(n_F \rightarrow n_F+m)$
transition. Below we shall derive an analytical expression for this
discontinuity for strictly vertical transition.

The case of half-filling is a quite special situation but actually
exhibits the general discontinuous behavior of the pair excitation in
modulated systems. If the last subband is less than half-filled, the
discontinuity in the imaginary part of the dielectric function will
appear for frequencies larger than $m \omega_c$.  If the last sub-band
is more than half-filled, the discontinuity in the imaginary part of
the dielectric function will appear for frequencies lower than $m
\omega_c$.  In this way the imaginary part of the dielectric function
also provides direct information about the filling of the last
subband. For integer fillings, such discontinuity will disappear. For
non-integer fillings but non-vertical excitations ($q_y \neq 0$), such
discontinuity will still be present.

The dielectric function within the random-phase approximation is
\begin{eqnarray}
  \epsilon(q_x,\omega) & = & 1 - 2 \pi r_s \frac{k_F}{q_x} \hbar
  \omega_c \times \nonumber \\ & & \sum_{n \neq n^{\prime}}
  \sum_{x_0}C_{nn^{\prime}}
  \frac{f_{n,x_0}-f_{n^{\prime},x_0}}{E_{n,x_0}-E_{n^{\prime},x_0} +
    \hbar (\omega + \imath \delta)} \, ,
\end{eqnarray}
where $r_s$ is the plasma parameters, $f_{n,x_0}$ is the Fermi
occupation function.  The coefficient $C_{nn^{\prime}}$
($n>n^{\prime}$) is given as
\begin{equation}
  C_{nn^{\prime}} = \frac{n^{\prime}!}{n!} X^{n-n^{\prime}}
  \Bigl[ L_{n^{\prime}}^{n-n^{\prime}}(X) \Bigr]^2 \exp{(-X)}
\end{equation}
with $X=(q_x l)^2/2$ and $L_n^m(X)$ is an associated Laguerre
polynomial.

For positive frequencies the imaginary part of the dielectric
function is\cite{Stewart95:R17036}
\begin{eqnarray}
& &   \text{Im}[\epsilon(q,\omega)]  = -2 \pi^2 r_s (k_F/q) \hbar
  \omega_c \times \nonumber \\ & & \sum_{m=1}^{\infty}
  \sum_{m^{\prime}=0}^{n_F} C_{m+m^{\prime},m^{\prime}}
  (f_{m+m^{\prime},x_j}-f_{m^{\prime},x_j}) Q^{mm^{\prime}} \, ,
  \label{imag}
\end{eqnarray}
where the function 
\begin{equation}
  Q^{mm^{\prime}} = \frac{\theta(|U_{m+m^{\prime}}-U_{m^{\prime}}
    - |m \hbar \omega_c - \hbar
    \omega|)}{\sqrt{(U_{m+m^{\prime}}-U_{m^{\prime}})^2 - (m \hbar
      \omega_c - \hbar \omega)^2}} 
\end{equation}
gives the square-root singularities.  In (\ref{imag}) we also sum
over the simple roots $x_j$ given by
\begin{equation}
  \cos{K x_j} =  - \frac{m \hbar \omega_c - \hbar
    \omega}{U_{m+m^{\prime}}-U_{m^{\prime}}} \, .
  \label{root}
\end{equation}
For small $(m \hbar \omega_c - \hbar \omega)/(U_{m+m^{\prime}} -
U_{m^{\prime}})$ we see that in the interval $0 < K x_0 < 2 \pi$ the
solutions of (\ref{root}) are $Kx_0 = \pi/2 + (m \hbar \omega_c - \hbar
\omega)/(U_{m + m^{\prime}} - U_{m^{\prime}})$ and $K x_1 = 3 \pi/2 -
(m \hbar \omega_c - \hbar \omega)/(U_{m + m^{\prime}} -
U_{m^{\prime}})$.  We
introduce a small energy shift $\delta = \omega - 4 \omega_c$ and
find from (\ref{imag}) and (\ref{root}) that for $\delta = 0^-$
\begin{eqnarray}
  & & \text{Im}[ \epsilon(q,\delta = 0^-)] = 4 \pi^2 r_S
  (k_F/q) \hbar \omega_c \times \nonumber \\ & & \left[
    \frac{C_{5,1}}{|U_5 - U_1|} + \frac{C_{6,2}}{|U_6 - U_2|} +
    \frac{C_{7,3}}{|U_7-U_3|} + \frac{C_{8,4}}{|U_8-U_4|} \right] \, .
\end{eqnarray}
On the contrary for $\delta = 0^+$ we find
\begin{eqnarray}
  & & \text{Im}[ \epsilon(q,\delta = 0^+) ]= 4 \pi^2 r_s
  (k_F/q) \hbar \omega_c \times \nonumber \\ & &  \left[
    \frac{C_{6,2}}{|U_6 - U_2|} + \frac{C_{7,3}}{|U_7-U_3|} +
    \frac{C_{8,4}}{|U_8-U_4|} + \frac{C_{9,5}}{|U_9-U_5|} \right] \, .
\end{eqnarray}
Therefore 
\begin{eqnarray}
  \Delta \text{Im}[ \epsilon(q_x)] & = & \text{Im} [\epsilon(q,\delta
  = 0^+)] - \text{Im}[\epsilon(q,\delta = 0^-)] \nonumber \\ & = &4
  \pi^2 r_S (k_F/q) \hbar \omega_c\times \nonumber \\ & & \left[
    \frac{C_{5,1}}{|U_5 - U_1|} - \frac{C_{9,5}}{|U_9-U_5|} \right] \, .
\end{eqnarray}
Similarly the discontinuity around $\omega = 3 \omega_c$ is given as
$C_{5,2}/|U_{5}-U_{2}| - C_{8,5}/|U_8-U_5|$, and etc.  Since the first
order potential element $U_n$ is linearly proportional to the
modulation potential, we see that apart from numerical constants the
discontinuity in the imaginary part of the dielectric function is
determined directly by the strength of the modulation potential.

Note also that the imaginary part of the dielectric function should
contain $i$ sub-singularities around each frequency band even for
partially filled subands as long as $q_y=0$. For nonzero $q_y$, in
principle one may have $i+1$ subsingularities as stated in
Ref.[\onlinecite{Stewart95:R17036}] where $i$ is an integer and counts
the resonance frequency.

For our numerical evaluation of the dielectric function we have
employed the following parameters for a typical modulated
GaAs/Al$_x$Ga$_{1-x}$As heterostructure, $\kappa=13$, $r_s=0.73$,
$E_F=10$ meV, and $m^*=0.067 m_e$.  The amplitude and period of the
modulation potential are $V_0=$1 meV, and $a=300 $ nm, respectively.
The calculation was done at zero temperature with $q_x = 0.2 k_F$,
$q_y = 0$.  The numerical parameters are thus the same as used in Ref.
[\onlinecite{Stewart95:R17036}] except that we have a strictly
vertical transition $q_y=0$ instead of the small $q_y=1 \times 10^6 $
m$^{-1}$.

We show in Fig.\ (\ref{f:im4}) the imaginary part of the dielectric
function around $\omega = 4 \omega_c$.  The discontinuity around
$\omega= 4 \omega_c$ is clearly resolved together with the
four subsingularities on each side of the frequency band.

If exchange and correlation effect are neglected, the imaginary part
of the dielectric function is proportional to the cross section of
spin-density excitations which may be measured in Raman scattering
when the polarization of the incoming and scattered light are
perpendicular.\cite{Klein84} Within this approximation the
discontinuity step in the imaginary part of the dielectric function is
therefore directly measurable.

The real part of the dielectric function is
\begin{eqnarray}
  & & \text{Re}[\epsilon(q,\omega)] = 1 - 4 \pi^2 r_s \frac{k_F}{q}
  \hbar \omega_c \times \nonumber \\ & & \sum_{m=1}^{\infty}
  \sum_{m^{\prime}=0}^{n_F} C_{m+m^{\prime},m^{\prime}} \Bigl(
  I_1 - I_2 + I_3 - I_4 \Bigr) \, ,
\end{eqnarray}
where
\begin{mathletters}
  \begin{eqnarray}
    I_1 & = &I(m \hbar \omega_c + \hbar \omega,U_{m+m^{\prime}},E_F -
    E_{m+m^{\prime}},U_{m+m^{\prime}}) \\
    I_2 & = & I(m \hbar \omega_c +
    \hbar \omega,U_{m^{\prime}},E_F - E_{m^{\prime}},U_{m^{\prime}}) \\
    I_3 & = &I(m \hbar \omega_c - \hbar \omega,U_{m+m^{\prime}},E_F -
    E_{m+m^{\prime}},U_{m+m^{\prime}}) \\
    I_4 & = & I(m \hbar \omega_c -
    \hbar \omega,U_{m^{\prime}},E_F - E_{m^{\prime}},U_{m^{\prime}})
    \label{i4}
  \end{eqnarray}
\end{mathletters}
are given by the Cauchy principal value integral
\begin{equation}
  I(a,b,c,d) \equiv \frac{1}{2\pi} \int_0^{\infty} d\phi \theta(c-d
  \cos{\phi}) P \frac{1}{a+b \cos{\phi}} \, .
\end{equation}
The integral may be found by a complex contour integration.  For
$(c/d)^2>1$, which is the case for completely occupied or unoccupied
sub-bands, the integral is simply
\begin{equation}
I = \left\{ \begin{array}{cc}
    \theta(c) 2 \pi K & (a/b)^2 > 1 \\
    0                      & (a/b)^2 < 1
    \end{array} \right. \, ,
\end{equation}
where the factor $K$ is
\begin{equation}
  K = \frac{1}{2 \pi a \sqrt{|1-(b/a)^2|}} \, .
\end{equation}
For the partially filled sub-bands where $(c/d)^2 < 1$ we find
\begin{equation}
  I = K \times \left\{ \begin{array}{cc}
      2 \pi - 4 \arctan{\gamma} & (a/b)^2 > 1, d> 0 \\
      4 \arctan{\gamma}           & (a/b)^2 > 1, d< 0 \\
      -2 \ln{|\frac{1-\gamma}{1+\gamma}|} & (a/b)^2 < 1, d>0 \\
      2 \ln{|\frac{1-\gamma}{1+\gamma}|} & (a/b)^2 < 1, d<0
    \end{array} \right. \, ,
  \label{I}
\end{equation}
where
\begin{equation}
  \gamma = \frac{1-b/a}{\sqrt{|1-(b/a)^2|}} \sqrt{\frac{1-c/d}{1+c/d}} \, .
\end{equation}
For the partially filled subband $m^{\prime}=5$ (at filling
$\nu=5.5$), the third argument of (\ref{f:im4}) is zero so that the real
part of the dielectric function is logarithmic divergent around $\hbar
\omega = 4 \omega_c$ as can be seen from (\ref{I}).  In addition the
real part of the dielectric function has four subsingularities for
frequencies slightly larger than $4 \omega_c$ and four
subsingularities for frequencies slightly smaller than $4 \omega_c$.
The real part of the dielectric function is shown in Fig.\ (\ref{f:re4})
around $\omega=4 \omega_c$.

In a far infrared absorption or a Raman scattering experiment where
the polarizations of the incoming and scattered photon are parallel,
the charge-density excitations are measured.  That is the scattering
cross section which is proportional to
$-\text{Im}[1/\epsilon(q,\omega)]$.  This function has peaks when
$\text{Re}[\epsilon(q,\omega)]=0$. For an unmodulated system, the only
peaks in the scattering cross section are those due to the
magneto-plasmon excitation. For a modulated system, certain spectral
weight is shifted back to the energy corresponding to the
electron-hole pair excitation. The peaks due to electron-hole pair
excitation are rather sharp but finite even when disorders are
negligible.  Furthermore, we found that the cross section has an
addition sharp peak at the $\omega=n\omega_c$ due to a logarithmic
singularity in the real part of the dielectric function.  This is
depicted in Fig.\ (\ref{f:ram4}).

In conclusion we have shown that a partial filling of the last
Landau band may be detected in optical spectroscopy either in the
spin-density excitation spectra or in the charge-density excitation
spectra.  The spectra will provide information about the modulation
potential and the magnitude of the partial filling.

\acknowledgements{We would like to thank R.R. Gerhardts and S. M.
  Stewart for interesting discussions.}

\newpage

\begin{figure}
\caption{Single-particle energies (lines) around the Fermi energy
  (broken line) as a function of the center coordinate, $Kx_0$, when
  the last subband is half-filled.  Electron-hole pair excitations
  with energies $\omega < \omega_c$, $\omega = \omega_c$ and $\omega >
  \omega_c$ are shown.}
\label{f:dis}
\end{figure}

\begin{figure}
  \caption{Imaginary part of the dielectric function,$\text{Im}[\epsilon(q_x,\omega)]$ as a function of $\omega/\omega_c$ for $\nu=5.5$, $V_0$=1 meV.}
\label{f:im4}
\end{figure}

\begin{figure}
  \caption{Real part of the dielectric function,$\text{Re}[\epsilon(q_x,\omega)]$ as a function of $\omega/\omega_c$ for $\nu=5.5$, $V_0$=1 meV.}
\label{f:re4}
\end{figure}

\begin{figure}
  \caption{$-\text{Im}[1/\epsilon(q_x,\omega)]$ as a function of $\omega/\omega_c$ for $\nu=5.5$, $V_0$=1 meV.}
\label{f:ram4}
\end{figure}


\begin{thebibliography}{10}

\bibitem{Weiss89:179}
D. Weiss, K. v.~Klitzing, K. Ploog, and G. Weinmann, Europhys. Lett. {\bf 8},
  179  (1989).

\bibitem{Winkler89:1177}
R.~W. Winkler, J.~P. Kotthaus, and K. Ploog, Phys. Rev. Lett. {\bf 62},  1177
  (1989).

\bibitem{Gerhardts89:1173}
R.~R. Gerhardts, D. Weiss, and K. v.~Klitzing, Phys. Rev. Lett. {\bf 62},  1173
   (1989).

\bibitem{Weiss89:13020}
D. Weiss {\it et~al.}, Phys. Rev. B {\bf 39},  13020  (1989).

\bibitem{Pfannkuche92:12606}
D. Pfannkuche and R.~R. Gerhardts, Phys. Rev. B {\bf 46},  12606  (1992).

\bibitem{Gerhardts90:1473}
R.~R. Gerhardts and C. Zhang, Phys. Rev. Lett. {\bf 64},  1473  (1990).

\bibitem{Zhang90:12850}
C. Zhang and R.~R. Gerhardts, Phys. Rev. B {\bf 41},  12850  (1990).

\bibitem{Cui89:2598}
H.~L. Cui, V. Fessatidis, and N.~J.~M. Horing, Phys. Rev. Lett. {\bf 63},  2598
   (1989).

\bibitem{Zhang90:2207}
C. Zhang, Phys. Rev. Lett. {\bf 65},  2207  (1990).

\bibitem{Manolescu95:1703}
A. Manolescu and R.~R. Gerhardts, Phys. Rev. B. {\bf 51},  1703  (1995).

\bibitem{Weiss93:4118}
D. Weiss {\it et~al.}, Phys. Rev. Lett. {\bf 70},  4118  (1993).

\bibitem{Rotter96:4452}
P. Rotter, M. Suhrke, and U. R\"{o}ssler, Phys. Rev. B {\bf 54},  4452  (1996).

\bibitem{Schlosser96:683}
T. Schl\"{o}sser, K. Ensslin, J.~P. Kotthaus, and M. Holland, Europhys. Lett.
  {\bf 33},  683  (1996).

\bibitem{Stewart95:R17036}
S.~M. Stewart and C. Zhang, Phys. Rev. B {\bf 52},  R17036  (1995).

\bibitem{Klein84}
M.~V. Klein,  in {\em Light Scattering in Solids I, Topics Appl. Phys. Vol. 8},
  edited by M. Cardona (Springer, Berlin, 1984).

\end{thebibliography}
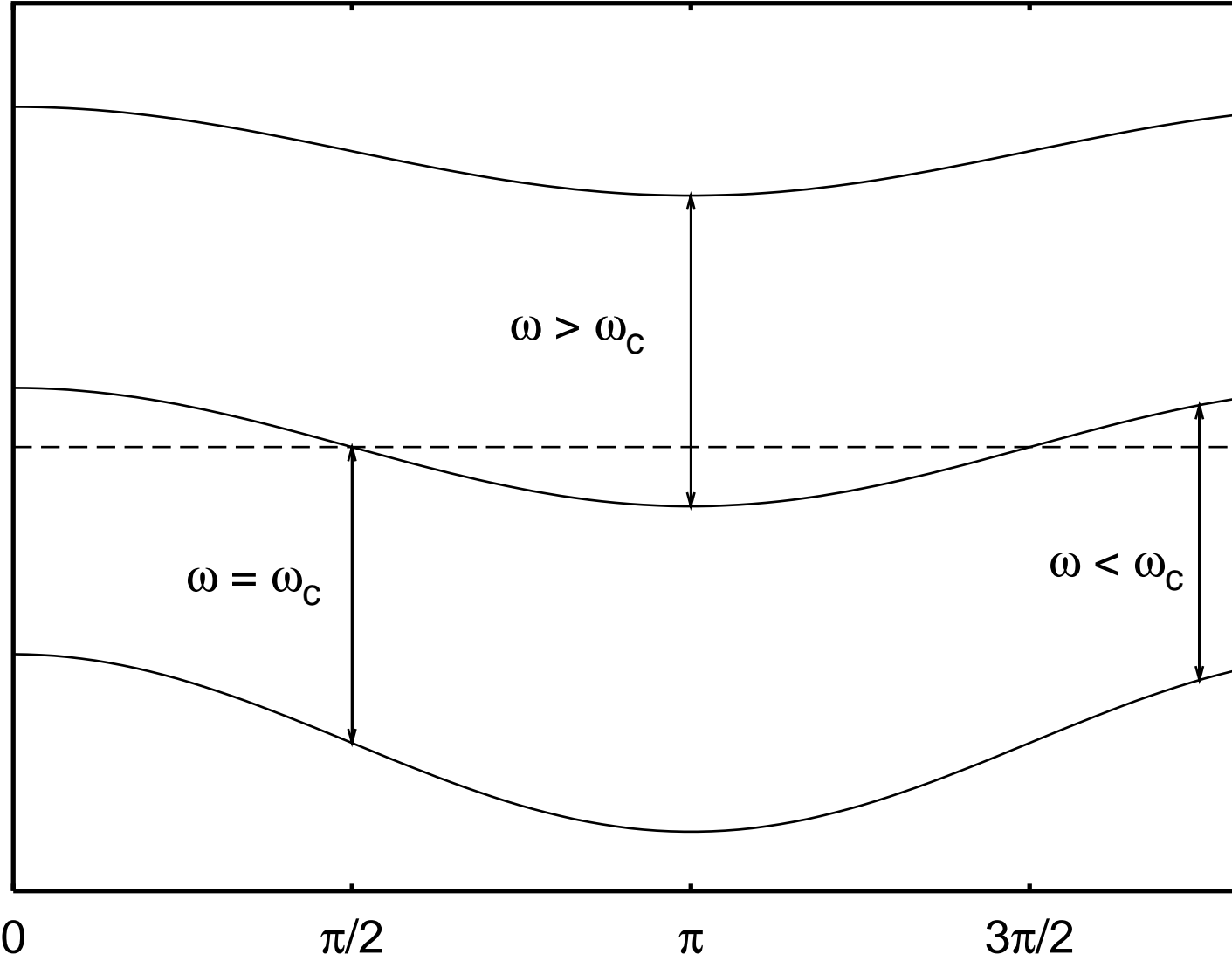
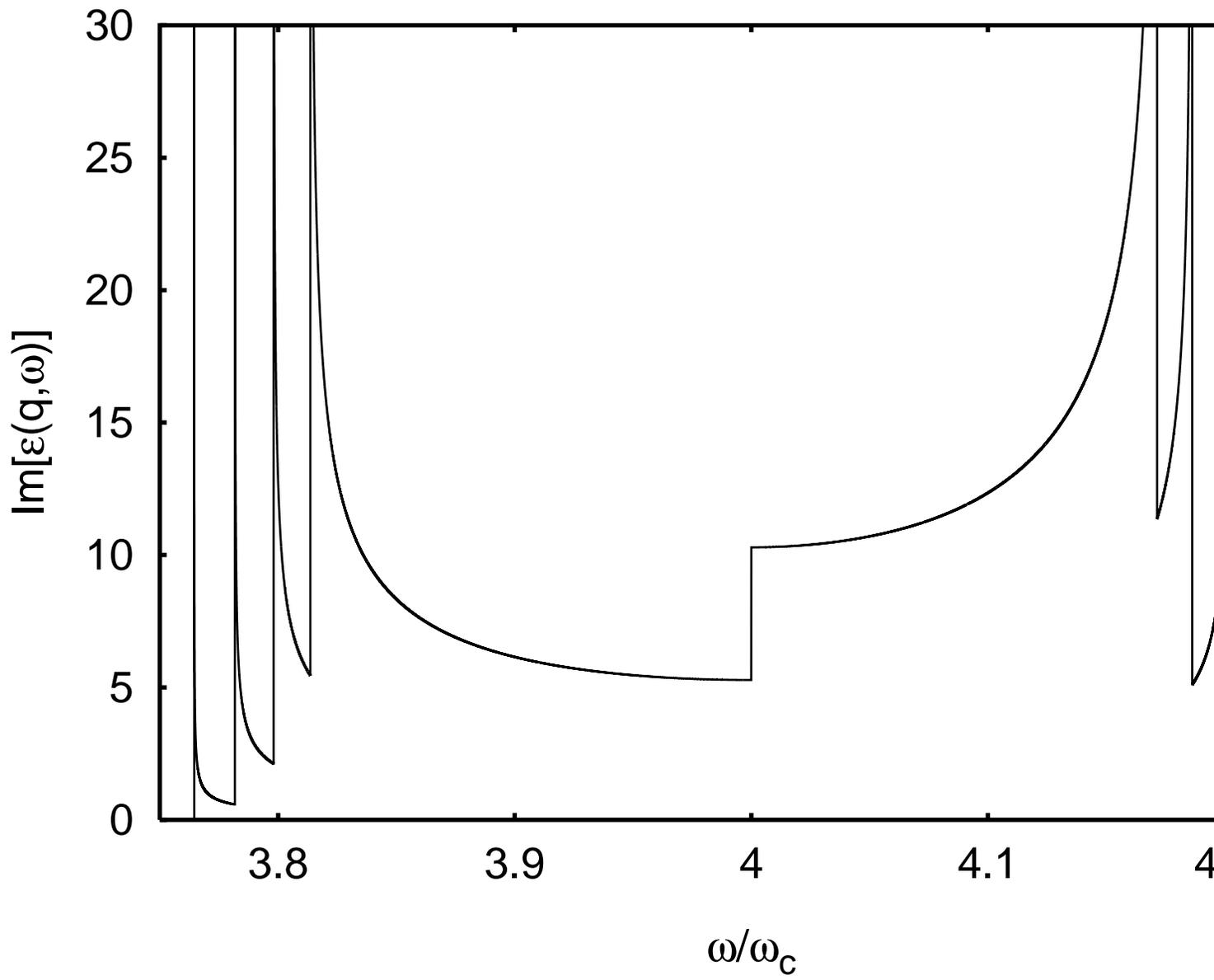
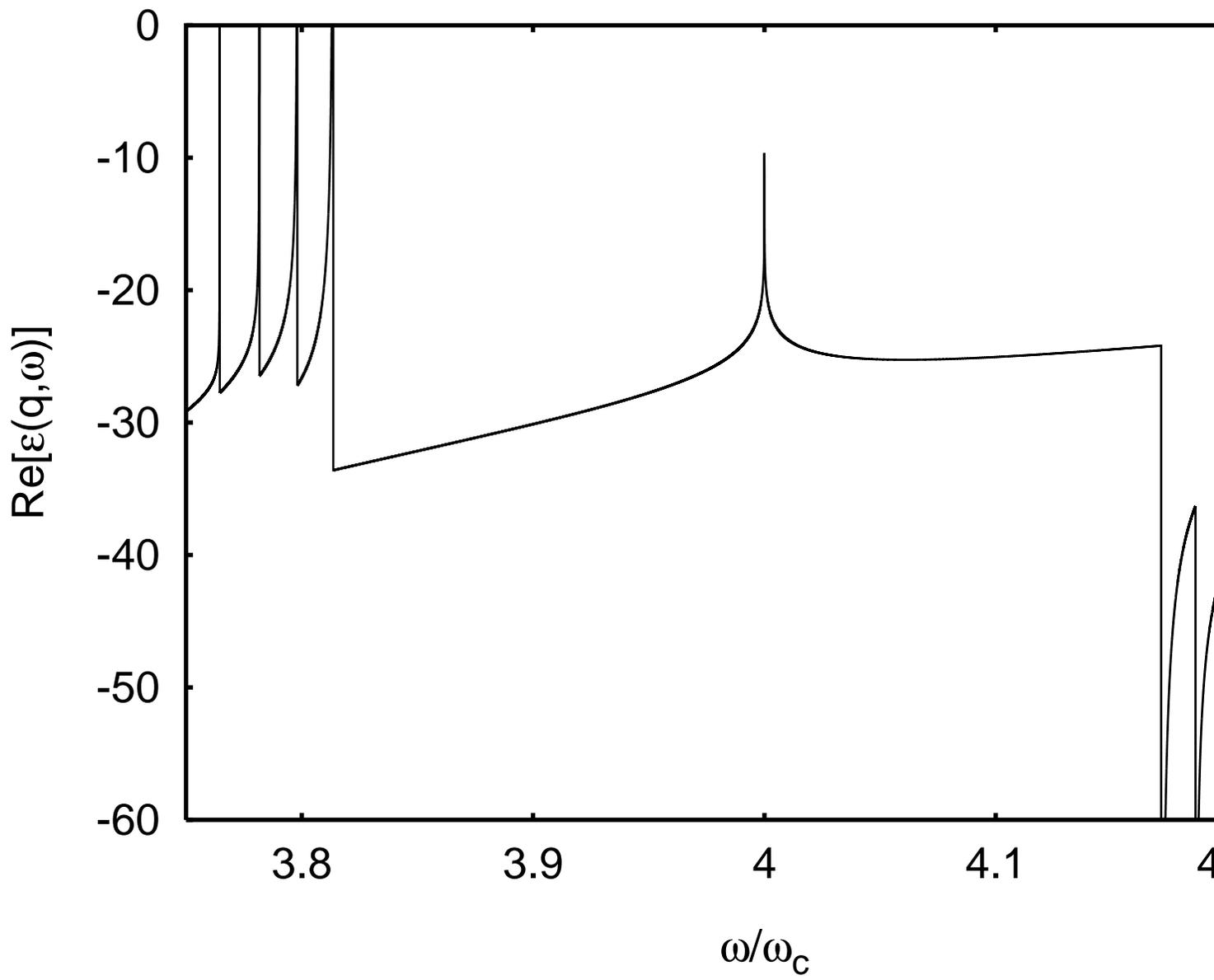
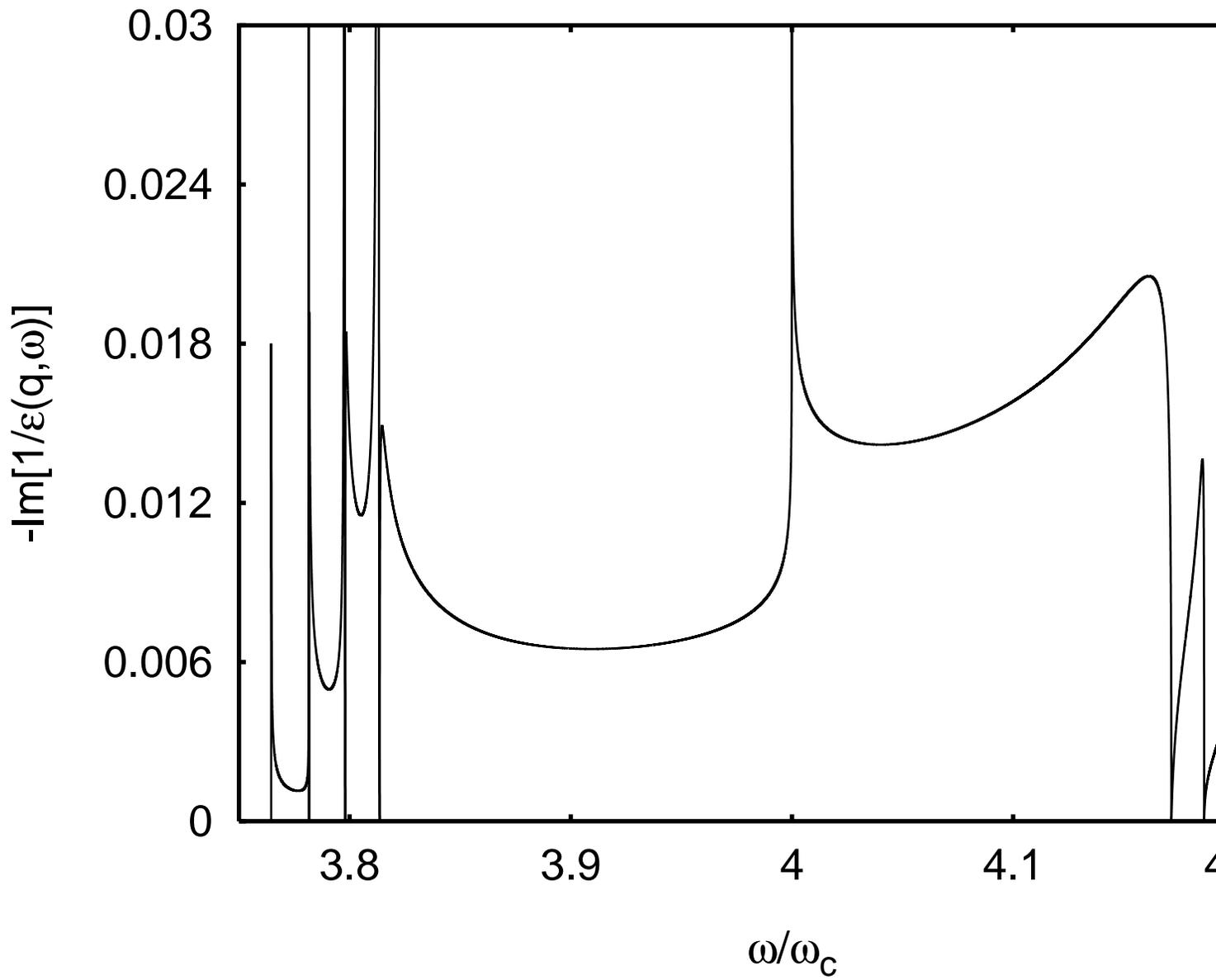
\end{document}